\documentclass[conference]{IEEEtran}
\IEEEoverridecommandlockouts
\usepackage{cite}
\usepackage{amsmath,amssymb,amsfonts}
\usepackage{algorithm}
\usepackage{algorithmicx}
\usepackage{algpseudocode}
\usepackage{footnote}
\usepackage{graphicx}
\usepackage{textcomp}
\usepackage{xcolor}
\usepackage{url}
\usepackage{float}
\usepackage{subfig}
\usepackage{caption}
\usepackage{diagbox}
\usepackage{array}
\usepackage{booktabs}
\usepackage{hyperref}
\usepackage{multirow}
\usepackage{float}
\usepackage{bm}
\usepackage[misc]{ifsym} 
\def\tablename{Table}
\def\BibTeX{{\rm B\kern-.05em{\sc i\kern-.025em b}\kern-.08em
    T\kern-.1667em\lower.7ex\hbox{E}\kern-.125emX}}
\begin{document}
\title{A Fast-Detection and Fault-Correction Algorithm against Persistent Fault Attack}

\author{\IEEEauthorblockN{Yukun Cheng$^{\textrm{\Letter}}$, Mengce Zheng, Fan Huang, Jiajia Zhang, Honggang Hu$^{\textrm{\Letter}}$ and Nenghai Yu}
\IEEEauthorblockA{\textit{Key Laboratory of Electromagnetic Space Information, CAS} \\
\textit{University of Science and Technology of China}, Hefei, China \\
Email: kuin@mail.ustc.edu.cn, hghu2005@ustc.edu.cn}
}

\maketitle

\begin{abstract}
	\emph{Persistent Fault Attack} (PFA) is a recently proposed \emph{Fault Attack} (FA) method in CHES 
	2018. It is able to  recover full AES secret key in the Single-Byte-Fault scenario. 
	It is demonstrated that classical FA countermeasures, such as 
	\emph{Dual Modular Redundancy} (DMR) and mask protection, are unable to thwart PFA.
	In this paper, we propose a fast-detection and fault-correction algorithm to prevent PFA. We  
	construct a fixed input and output pair to detect faults rapidly. Then we build two extra redundant tables to store the relationship between the adjacent elements in the S-box,
	by which the algorithm can correct the faulty elements in the S-box. 
	Our experimental results show that our algorithm can  effectively prevent PFA in both  Single-Byte-Fault and  Multiple-Bytes-Faults scenarios. Compared with the classical FA countermeasures, our algorithm has a much better effect against PFA.
	Further, the time cost of our algorithm is 40\% lower than the classical FA countermeasures.
\end{abstract}

\begin{IEEEkeywords}
	Fault Attack Countermeasure, AES, Fault Correction,  Persistent Fault Attack
\end{IEEEkeywords}

\section{Introduction}
\label{sec1}

Traditional cryptographic algorithms focus on hard mathematical problems. However, to provide a reliable and convenient computing environment, cryptographic algorithms are implemented in embedded cryptographic devices nowadays.  Implementation-based attacks bring a severe threat to embedded cryptographic devices. 
Among them, \emph{Fault Attack} (FA) first proposed by Habing \cite{4323904} in 1965 is a powerful one with strong positivity and flexibility.
Since Boneh et al.   
made use of FA to break the RSA-CRT \cite{10.1007/3-540-69053-0_4} in 1997, 
it has received intensive attention in the area of cryptanalysis. Recently, 
a large number of cryptographic algorithms have been found vulnerable to FA, such as ECC \cite{biehl2000differential}, 
AES \cite{piret2003differential}, PRESENT \cite{wang2010differential} and 
LED \cite{jovanovic2012fault}. The progress of FA mainly includes 
two branches. On one hand,
the technique of fault injection has advanced continuously to alter either the control flow or the internal state of the cryptographic algorithm. On the other hand, a large body of 
related work on fault analysis has been conducted to develop novel methods to recover the key used in the encryption.

Fault injection aims to induce cryptographic device faults using some physical methods, 
  such as lowering support voltage \cite{biehl2000differential}, clock glitch \cite{amiel2006fault}, 
  electromagnetic injection \cite{schmidt2007optical}, and temperature variation \cite{govindavajhala2003using}. 
  All the above methods are low-cost but low-precision. The adversary needs to collect a massive number
  of ciphertexts and sift out the useful ones. Moreover, if the adversary is more powerful, laser \cite{skorobogatov2005semi} or \emph{Focused Ion Beam} (FIB) \cite{torrance2009state} 
  are more suitable for fault injection due to their higher precision. The adversary 
  can affect considerable information of the cryptographic implementation, no matter whether the 
  hardware alteration protection exists or not.  Skorobogatov et al. demonstrated that the
  \emph{Static Random-Access Memory} (SRAM) \cite{skorobogatov2002optical} and registers \cite{courbon2014adjusting} of the microcontroller ($\mu$C) are vulnerable to 
  laser. In addition to $\mu$C, ASIC \cite{selmane2008practical} and FPGA \cite{canivet2011glitch} are also easily affected by laser injection.

  Following the fault injection phase, the adversary needs to exploit the faulty ciphertexts to derive the key used in the encryption process.
  Some fault analysis methods, such as  \emph{Differential Fault Analysis} (DFA) 
  \cite{piret2003differential} \cite{biham1997differential},  focus on the differences between the faulty and the correct ciphertext pairs.
  Some alternatives, e.g. \emph{Statistical Fault Analysis} (SFA) \cite{fuhr2013fault}, only focus on the statistical properties of faulty ciphertexts. Besides, Clavier showed that
  ineffective faults can be employed to obtain security information via 
  \emph{Ineffective Fault Analysis} (IFA) \cite{clavier2007secret} in CHES 2007.  More recently, 
  \emph{Statistical Ineffective Fault Analysis} (SIFA) \cite{dobraunig2018sifa}, a combination of IFA and SFA, was proposed 
  as a novel method in CHES 2018.

  The countermeasures against FA aim to detect faults or prevent the use of faults at two primary levels --- both physical level and algorithm level.
  The physical level countermeasures aim to detect faults with physical components like the ring
  oscillators \cite{bohl2012fault} \cite{deshpande2016configurable} or supply voltage supervisors \cite{zussa2014efficiency}. These physical  level countermeasures with high-precision can detect faults rapidly.
  However, they need to be deployed on the circuits at the design stage. This drawback leads to the low flexibility of the device when new vulnerabilities come to light. The algorithm level countermeasures tend to prevent FA by adding redundant information, such as \emph{Dual Modular Redundancy} (DMR) \cite{bar2006sorcerer}, \emph{Bytes Scrambling} (BS) \cite{joye2007strengthening} and error detection code \cite{karpovsky2004robust}. These algorithm countermeasures can be applied to different ciphers flexibly. However, the user needs to afford higher cost and detection delay.

  In CHES 2018, Zhang et al.  proposed the concept of \emph{Persistent Fault Attack} (PFA) \cite{zhang2018persistent}. And in CHES 2020, they demonstrated the feasibility of PFA in  practical applications \cite{zhang2020persistent}. Unlike other
  FA methods, the fault model of PFA assumes that the fault in the cryptographic device is persistent 
  rather than transient or permanent. The fault can persist until the device reboots.
  Under this condition, PFA demonstrates its advantage in the FA research area. In the Single-Byte-Fault scenario, it can recover the full AES secret key using less than 2000 ciphertexts. It can also be applied in the Multiple-Bytes-Faults scenario to reduce the number of key guesses.

  Besides, some works showed that PFA is robust against the classical FA countermeasures.
  In \cite{zhang2018persistent},  the authors showed that PFA can bypass the DMR.
  Moreover, in \cite{pan2019one}, the higher-order masked S-box was verified to be ineffective for PFA. Although some physical level countermeasures, e.g. the ring oscillators, can be deployed on the top of sensitive regions to detect faults, there are few algorithm level countermeasures that can prevent PFA effectively.

\subsection{Contributions}
\label{subsec1.2}
In this paper, we propose a novel algorithm level countermeasure against PFA using a fast-detection and fault-correction algorithm on AES.
The main contributions are summarized as follows:
\begin{enumerate}
    \item We propose a novel algorithm that can detect and correct constant faults for all elements in the S-box to keep the correctness of ciphertexts.  The effectiveness of our algorithm is evaluated in both Single-Byte-Fault and Multiple-Bytes-Faults scenarios and compared with two classical FA countermeasures --- DMR and BS.
    We demonstrate that our algorithm has less time cost than the two countermeasures.
    The experimental results show that our algorithm has a better protection effect against PFA.
    \item To ensure the practicability and correctness of our algorithm, we conduct experiments with AES in both software implementation in $\mu$C and hardware
          implementation in FPGA. 
          In the FPGA implementation, our algorithm is improved using the pre-correction mode
          for cost reduction.
    \item Finally,  some problems are discussed besides our algorithm itself. First, the discussion is about the relationship between the performance of PFA and the number of faults.
    Then we compare our algorithm with \emph{Triple Modular Redundancy} (TMR) which has a similar majority voting calculation.
    In the end, we discuss the expansibility of our algorithm.  
\end{enumerate}

\subsection{Organizations}
\label{subsec1.3}

The rest of the paper is organized as follows: Sect. \ref{sec2} introduces the background.
Sect. \ref{sec3} highlights the core idea, the process of our algorithm, and the comparison of  the time cost of our algorithm
with other FA countermeasures. Sect. \ref{sec4} discusses the effectiveness of our algorithm in Multiple-Bytes-Faults scenario.
Sect. \ref{sec5} shows the experimental results in both software and hardware implementations.  The discussions are provided in Sect. \ref{sec6}.
Finally, we conclude the paper in Sect. \ref{sec7}. 

\section{Preliminaries}
\label{sec2}

\subsection{AES Algorithm and Implementation}
\label{subsec2.1}

In October 2000, the Rijndael algorithm was chosen as the \emph{Advanced Encryption Standard} 
(AES) \cite{dworkin2001advanced}. Afterward, AES is widely used to protect the information 
confidentiality of embedded devices. AES uses a key with the length of 128 bits, 192 bits or 256 bits to encrypt 
data with a block size of 128 bits. Because PFA in \cite{zhang2018persistent} was developed on the basis of AES-128, our algorithm
also focuses on AES-128 in this paper.

AES consists of $r$ transformation rounds. The value of $r$ depends on the key length. In AES-128, $r$
is set to 10.
AES operates on a column-major order matrix of bytes with the size of $4 \times 4$, termed the \emph{State}.
The round functions include four byte-oriented operations: AddRoundKey, SubBytes, ShiftRows and MixColumns.  Especially in the last round, there is no MixColumns
operation according to the specification. Moreover, an extra KeyExpansions operation is used to generate the round key.

For the sake of convenience and efficiency, the nonlinear SubBytes operation is usually replaced by the look-up table in practical implementation. The look-up table prestores
the calculation results as constant in the registers
or SRAM of the device which are vulnerable to FA. 

In \cite{zhang2018persistent}, the linear operation, ShiftRows, was removed for the simpler analysis. Because the ShiftRows operation
only changes the sequence of the bytes in the state, rather than the value of the bytes in the state. In this paper, we also use this simplification.

\subsection{Countermeasures against Fault Attack}
In this paper, we pay more attention to the algorithm design. DMR and BS are introduced as two classical algorithm level countermeasures.
  We review the two countermeasures below.
\label{subsec2.2}

%$\emph{\textbf{Dual Modular Redundancy}
$\emph{\textbf {Dual\ Modular\ Redundancy.}}$
DMR is a fault detection technique to detect
errors in the circuits directly. Bar-El et al. applied it to prevent FA in \cite{bar2006sorcerer}.
This countermeasure is readily adopted in commercial solutions due to its reliability and security. 

\begin{figure*}[tb]
	\centering	
	\subfloat[REDMR]
	{
		\begin{minipage}[t]{0.45\textwidth}
			\centering   
			\includegraphics[width=0.8\textwidth]{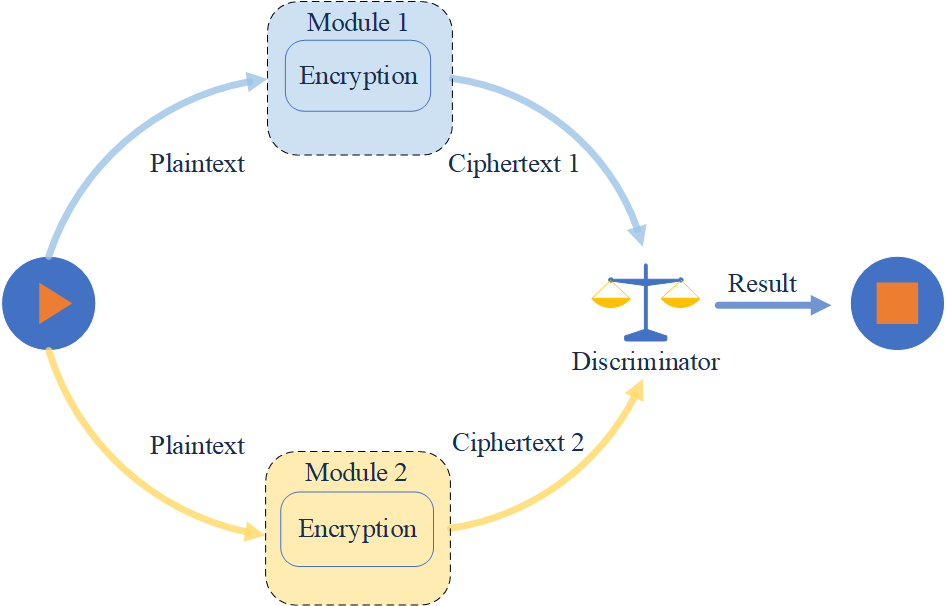} 
		\end{minipage}
	}
	\subfloat[IDDMR]
	{
		\begin{minipage}[t]{0.45\textwidth}
			\centering 
			\includegraphics[width=0.8\textwidth]{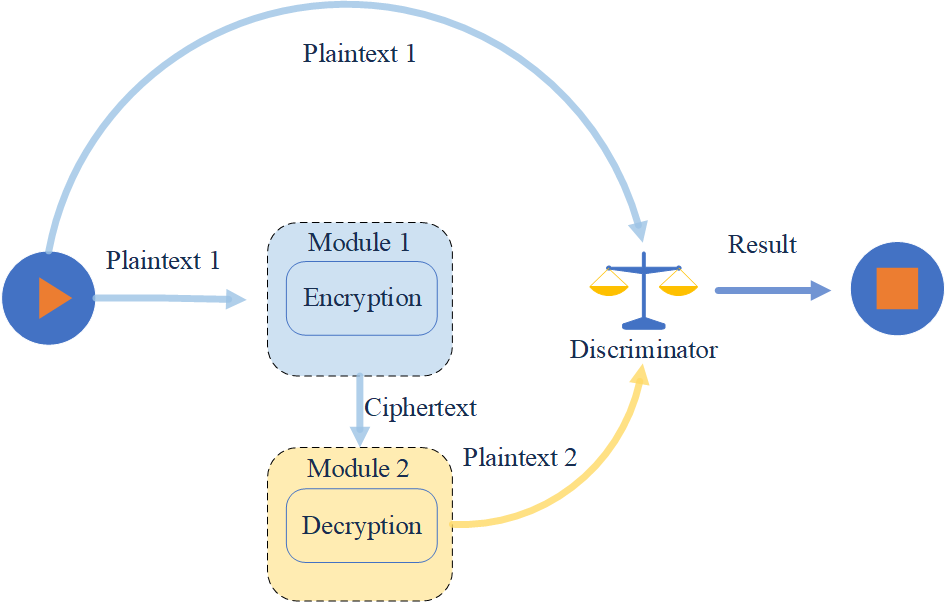} 
		\end{minipage}
	}	
	\caption{Dual Modular Redundancy scheme.}
	\label{fig1}
\end{figure*}

As shown in Fig.    \ref{fig1}, the DMR scheme uses two modules to perform cryptographic operations. It is called as  \emph{Redundant Encryption based DMR} (REDMR)  if module 1 and module 2 both perform encryption operation. The
discriminator of REDMR compares the ciphertexts of two modules in order to detect faults.  In contrast,  
if module 1
performs encryption operation, and module 2 performs decryption operation with the ciphertext of module 1, 
this situation is named as \emph{Inversive Decryption based DMR} (IDDMR). The
discriminator of IDDMR compares the original plaintext with the plaintext from module 2 to detect faults.
Based on the result of discriminator, some defenses can be deployed, such as \emph{No Ciphertext Output} (NCO), 
\emph{Zero Value Output} (ZCO) and \emph{Random Ciphertext Output} (RCO).  Cryptographic devices are forbidden to output ciphertexts with NCO defense. With ZCO defense, the ciphertexts are only composed of zero. And cryptographic device outputs random values as the ciphertexts with RCO defense.

\begin{figure}[tb]
  \centering
  \includegraphics[width=0.3\textwidth]{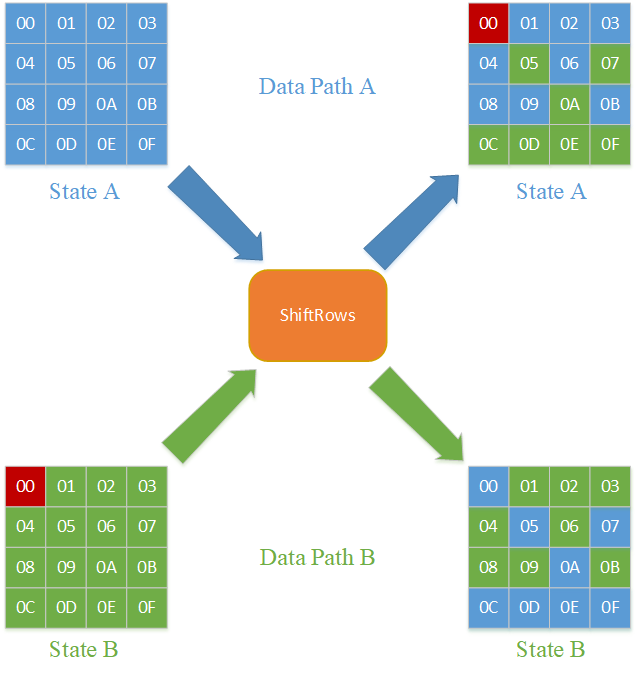}
  \caption{Scrambling bytes with one fault in AES state.}
  \label{fig2}
\end{figure}

$\emph{\textbf{Bytes Scrambling.}}$
In addition to detection techniques, another effective countermeasure is to make it hard to
make use of faults for the adversary. BS cuts faults out from ciphertexts to shadow the information of 
faults. Compared with DMR, the advantage of BS is that it has no discriminators and thus cannot be the target of high order FA \cite{kim2007fault}.

In \cite{joye2007strengthening}, Joye et al. first applied BS to block cipher and proposed a strengthening hardware AES implementation
against FA. The main idea is to use two modules to perform two encryptions in the parallel data path and scramble bytes in the state between the two executions.
The ShiftRows operation in AES can be used to perform this scrambling. Note that the adversary is able to inject faults into only one module and observe the ciphertext of this module. We assume that one fault has been injected into the $0$-th byte of the state B before the ShiftRows operation in the last round. As illustrated in Fig.    \ref{fig2}, we can see that the half state bytes on data path A (the blue elements) 
are swapped with their corresponding state bytes on data path B (the green elements) using the special ShiftRows operation. If there are no faults in the two states, the special ShiftRows operation has the same effect as the original ShiftRows operation. However, if there is one fault in state B (the red element), the fault is moved to state A following the ShiftRows operation. As we assumed, the ciphertexts that the adversary observes are correct and any fault analysis of state B is ineffective.

\subsection{Persistent Fault Attack}
\label{sec2.3}
As a novel FA method, PFA focuses on the fault which remains between transient and permanent, called persistent
fault. In \cite{zhang2018persistent}, the fault model assumed the affected constant in AES S-box stays faulty unless refreshed, which means the fault can affect several consecutive encryptions.

The core idea of PFA is to perform a statistical analysis of the distribution of each ciphertext candidate. For instance, AES has 256 elements in the S-box. For the $j$-th byte plaintext input, the number of ciphertext candidates $c_j$ is $2^8$.  Because of the avalanche effect, for each ciphertext candidate, the distribution probability
$Pr(c_j)$ is near to $2^{-8}$ when the encryption is correct. For simplification, we ignore the ShiftRows operation because it is a linear process. The ciphertext $c_j$ is composed of the $10^{th}$ round S-box output $y_j$
and the last round key $k_j$ using bitwise xor, which is equivalent to the following:
\begin{equation}
  \label{eq1}
  c_j = y_j \oplus k_j 
\end{equation}

For example, suppose that one fault tampers with the element $v$ in AES S-box to another element $v^*$. It leads to the lack of one candidate in the S-box output $y_j$. The distribution probability of S-box output $Pr(y_j)$ can be described as follows:
\begin{equation}
  \label{eq2}
  Pr(y_j) = 
  \begin{cases}
    0, & y_j = S[v] \\
    2 \times 2^{-8}, & y_j = S[v^*] \\
    2^{-8},& otherwise \\
  \end{cases}
\end{equation}

According to Eq.(\ref{eq1}) and Eq.(\ref{eq2}), the distribution probability of each ciphertext candidate $Pr(c_j)$  is identical to the S-box output $Pr(y_j)$ because $k_j$ is fixed, which can be easily represented as:
\begin{equation}
  \label{eq3}
  Pr(c_j) = Pr(y_j \oplus k_j) =
  \begin{cases}
    0, & c_j = S[v] \oplus k_j\\
    2 \times 2^{-8}, & c_j = S[v^*] \oplus k_j\\
    2^{-8},& otherwise \\
  \end{cases}
\end{equation}

The process of PFA against the $j^{th}$ byte key $k_j$ is described as follows. First, the adversary injects persistent faults into the cryptographic device using the laser or 
other physical methods. Then the adversary encrypts with random plaintexts repeatedly and collects the ciphertexts. With a sufficient number of ciphertexts (generally needs thousands of ciphertexts),
the adversary can perform a statistical analysis of the distribution of each ciphertext candidate. According to the result of the theoretical analysis, 
the candidate with the most number of occurrences $c^{max}_j$ and the candidate with the least number of occurrences $c^{min}_j$ can be searched to recover the last 
round key $k_j$ using a simple bubble sort.

It is interesting that every ciphertext candidate is helpful for recovering the last round key $k_j$, whether it is the most probable, the least probable, or another ciphertext byte value. 
It is clear that the adversary can recover $k_j$ with $c^{max}_j$ and $c^{min}_j$ according to  Eq.(\ref{eq4}) and Eq.(\ref{eq5}):
\begin{equation}
  \label{eq4}
  k_j = c^{min}_j \oplus S[v] 
\end{equation}

\begin{equation}
  \label{eq5}
  k_j = c^{max}_j \oplus S[v^*] 
\end{equation}

With other ciphertext candidates $c^{others}_j$, the adversary is unable to recover $k_j$ directly. However, the adversary can use these candidates to eliminate impossible candidates of
$k_j$, which can be written as:
\begin{equation}
  \label{eq6}
  k_j \neq c^{others}_j \oplus S[v] 
\end{equation}

Actually, in \cite{zhang2018persistent}, the authors pointed out that the statistical analysis may produce faulty positives of $c^{max}_j$ and $c^{min}_j$ (i.e. one of $c^{others}_j$ is mistaken for $c^{max}_j$ or $c^{min}_j$)
because $c^{max}_j$ and $c^{min}_j$ can be obviously distinguished from others only when the total number of ciphertexts $n$ is
sufficient. Therefore, eliminating the impossible candidates may be more effective with a small number of ciphertexts $n$.

It is worth noting that, in practice, the adversary has no prior knowledge of the value of the faulty element in the S-box. In other words,
the adversary is unable to distinguish $v$ or $v^*$ from all elements in the S-box. However, it does not mean that PFA is impractical.
In \cite{zhang2020persistent}, Zhang et al. solved this problem by adding an extra step to search $v$ and $v^*$ before the recovery step.
The experimental result demonstrated that the number of ciphertexts to search $v$ and $v^*$ is less than the number of ciphertexts to recover $k_j$. There is no need to collect extra ciphertexts
for the search step.

\section{A Novel Algorithm against PFA}
\label{sec3}

In this section, we introduce a novel algorithm against PFA. First, we provide our fault model and core idea. After that, 
we describe the process of our algorithm and discuss some details like the time cost analysis and the comparison with other FA countermeasures.

All analysis in this section is aimed for the Single-Byte-Fault scenario. The more complicated Multiple-Bytes-Faults scenario is discussed in Sect. \ref{sec4}.

\subsection{Fault Model}

To prevent PFA, our fault model follows the original PFA fault model in \cite{zhang2018persistent}, which is as follows:
\begin{itemize}
  \item The adversary injects persistent faults into the cryptographic device before the encryption.
  \item The persistent faults randomly corrupt the stored constant (i.e. the S-box table in AES implementation) in the storage of the cryptographic device.  
  \item The persistent faults persist for some time until the storage is refreshed or the device is reboot. 
  \item The adversary is able to use the faulty device to encrypt random plaintexts and collect ciphertexts for subsequent analysis.
\end{itemize}

\subsection{Core Idea}
The core idea of our algorithm is to keep the correctness of all ciphertexts. As a matter of fact, the successful fault injection does not mean  that  the FA is successful. The information which the adversary can directly exploit to recover the key  exists in the faulty ciphertexts.
If the encryption does not access the faulty element in the  S-box, the round computation and the  ciphertexts will be correct. 
Such correct ciphertexts are useless for the adversary. If the adversary is unable to collect enough number of the faulty ciphertexts, PFA will be failed. 

To this end, our algorithm is composed of three stages. In the first stage, we propose a fast-detection mechanism to detect the existence of the fault in each element in the S-box. 
If the fault is detected, a fault-correction mechanism is set up to correct the fault and ensure the correctness of ciphertexts in the second stage. The fault-correction mechanism
makes use of the relationship between the adjacent elements in the S-box to build extra redundant tables. Using these redundant tables, the algorithm can correct the faulty elements in the S-box. In the final  stage, the cryptographic device encrypts with the plaintexts and 
produces the correct ciphertexts. The details of the two mechanisms are described below.

\subsection{The Fast-Detection Mechanism}
\label{sec3.3}
$\emph{\textbf{The Detection Mechanism Using the SubBytes Module.}}$
According to the fault model, the fault is injected into the storage so as to corrupt the elements in the S-box. A simple detection solution is to traverse all the elements in the S-box
to detect faults. Due to excessive overhead, we tend to utilize available resources more than a new detection module.
Based on the modular design, the four operations in the process of AES encryption (AddRoundKey, SubBytes, ShiftRows and MixColumns) can be divided into four 
standalone modules. Fortunately, the SubBytes module can be used to detect faults because it uses S-box to produce the outputs. 
Note that the output of the S-box is fixed with a fixed input. 
On account of the length of the plaintext is 16 bytes, the SubBytes module of AES can only access at most 16 elements in the S-box in each iteration. Therefore, the detection mechanism
can only detect at most 16 elements in the S-box in each iteration.

A trivial method is inputting all elements in the S-box into the SubBytes module and prestoring the correct outputs. The detection mechanism compares the calculations of the module with prestored values to detect the fault. However, this method requires to prestore $\frac{256}{16} = 16$ pairs of fixed inputs and outputs. And the comparison needs to be repeated 16 times. The cost of this method is excessive thus we need to find a new way to detect the fault more effectively.

$\emph{\textbf{Improvement with the Closed Loop.}}$
 To improve the detection mechanism, we tend to find one pair of fixed input and output that can detect all elements in the S-box. For this purpose,
  we introduce a concept termed closed loop. If the output of the SubBytes module after a few iterations is equal to the initial input, all elements used in the SubBytes module are in the same closed loop. Fig.    \ref{fig2_3} shows a closed loop in AES S-box with 59 elements.
 Note that different loops have different lengths. We make the following assumptions:

 \begin{figure}[tb]
  \centering
  \includegraphics[width=0.45\textwidth]{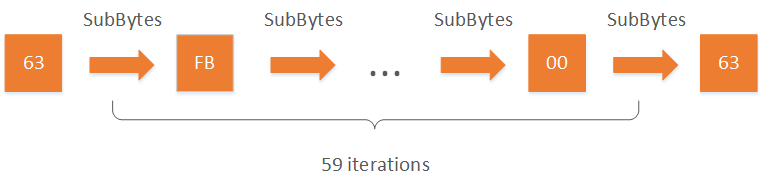}
  \caption{A closed loop in AES S-box with 59 elements.}
  \label{fig2_3}
\end{figure}
 
 \begin{itemize}
  \item There are $n$ elements in the S-box of the cryptographic algorithm.
  \item All elements in the S-box are assigned to $k$ closed loops.  The length of the $i$-th
  closed loop can be denoted as $l_i$.
  \item The length of the input of the SubBytes module is $m$. With a fixed input, the SubBytes module can access all elements in the S-box after $t$ iterations.
\end{itemize}

Therefore, the fixed input needs to be constructed skillfully. Let $d_i$ denote the number of bytes in the input which is distributed to the $i$-th closed loop. Let $r_i$ denote the minimal iteration times the SubBytes module requires to traverse all elements of the $i$-th closed loop,
which can be calculated as:

\begin{equation}
  \label{eq cl1}
  r_i = \lceil \frac{l_i}{d_i} \rceil,\qquad i=1,...,k
\end{equation}

According to the bucket theory, the iteration time $t$ depends on the maxinum of $r_i$, which can be described as:

\begin{equation}
  \label{eq cl2}
  t = max\{r_i\} = max\{\lceil \frac{l_i}{d_i} \rceil \},\qquad i=1,...,k
\end{equation}
 
To accelerate the detection mechanism, we need to minimize the iteration time $t$ with the constraint condition $\sum_{i=1}^{k}d_i = m$.
The best solution is to make the iteration of each closed loop near. Thus the set of $d_i$ can be calculated as:

\begin{equation}
  \label{eq cl3}
  \begin{aligned}
  \{d_i\} &= argmin(t) \\
  &= argmin( max\{\lceil \frac{l_i}{d_i} \rceil \}),\qquad i=1,...,k
  \end{aligned}
\end{equation}

$\emph{\textbf{The Fast-Detection Mechanism of AES S-box.}}$
Our detection method is described as follows:

\begin{itemize}
  \item[(1)] We construct one pair of fixed input $P$ and output $C$ which is processed in the SubBytes.  
  \item[(2)] We process the input in the SubBytes. After $t$ iterations, the SubBytes module can access all elements in the S-box and generate the output $C'$.
  \item[(3)] We compare $C$ with $C'$. If $C \neq C'$, the fault is detected.
\end{itemize}

\begin{figure}[tb]
  \centering
  \includegraphics[width=6cm]{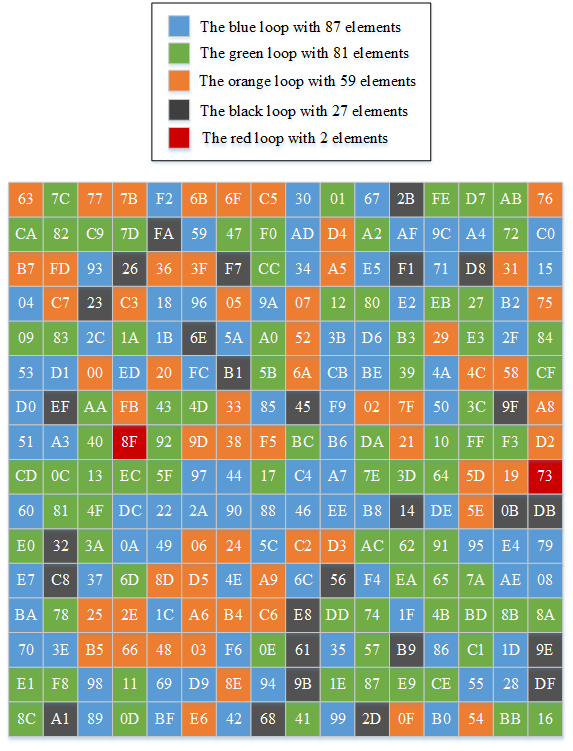}
  \caption{The elements of different loops in AES S-box.}
  \label{fig3}
\end{figure}

\begin{figure}[tb]
  \centering
  \includegraphics[width=8cm]{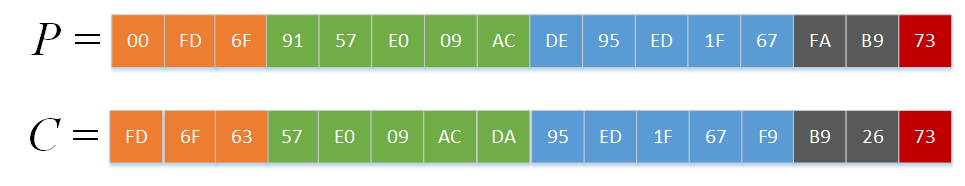}
  \caption{The fixed input $P$ and output $C$ constructed in the  fast-detection mechanism.}
  \label{fig3_4}
\end{figure}

 As shown in Fig.    \ref{fig3}, the elements of different loops in AES S-box are marked with different colors.
There are five closed loops in AES S-box. The longest loop (the blue loop) has 87 elements, and the shortest loop (the red loop) has only 2 elements.  
Thus,
the fixed input $P$ and output $C$ that constructed in our fast-detection mechanism are shown in Fig.    \ref{fig3_4}.

The minimum number of iterations of each loop is 20, 17, 18, 11, and 2, respectively. Consequently, we 
set the minimum number of iterations to 20 to guarantee that each element in the S-box is detected. Then the fixed output can be calculated and shown in Fig.    \ref{fig3_4} as well.
Because the number of iterations is minimum on the whole, the detection mechanism reaches the fastest speed in our approach.

However,the shortest loop is a special case. Note that the byte of this loop in both fixed input and output are the same. If a fault is able to alter the output of the element 73 to itself, the fixed input $P$ and output $C$ in Fig. \ref{fig3_4} is unable to detect this fault.  In fact, the adversary has no ability to inject targeted faults  under the current technology conditions. 
Even assuming that the adversary has this ability, we can calculate another  fixed output $\hat{C}$ after 21 iterations. With $C$ and $\hat{C}$, the two elements in the shortest loop can be detected meanwhile.

\subsection{The Fault-Correction Mechanism}
\label{sec3.4}
Once the fault is detected by the fast-detection mechanism, the fault-correction mechanism is designed to correct faults and ensure  the correctness of ciphertexts. 
As described in Sect. \ref{sec2}, two redundant modules are demonstrated that they can detect the fault. We plan to correct the fault with redundant modules as well. However, two redundant modules are unable to correct the fault because the discriminator in the DMR is unable to distinguish which output is correct.
If we use more than two redundant modules, the number of the correct outputs is more than the number of the faulty outputs. To guarantee the correctness of the output of the discriminator,
we can add a new rule that the minority should follow the majority.

An existing redundant resource is the adjacencies of each element in the S-box.  In the S-box based on look-up table, each element has four adjacencies in four directions (up, down, left and right). Thus the S-box is able to provide 4 times redundancy which is enough to correct the fault.
However, in the S-box, one element is hard to be calculated by its adjacencies directly.
 We need to build a relationship between the adjacent elements by the bitwise xor computation to achieve this goal.

In this subsection, we use $V_{i,j}$ to denote the element in AES S-box, which is in row $i$, column $j$. Respectively, we use 
$V_{i-1,j}$, $V_{i+1,j}$, $V_{i,j-1}$ and $V_{i,j+1}$ to denote four adjacencies of $V_{i,j}$. And the bitwise xor computation results 
between $V_{i,j}$ and its adjacencies can be denoted as $C_{up}$, $C_{down}$, $C_{left}$ and $C_{right}$, respectively. 
Take the up direction as an example, the relationship among $V_{i,j}$, $V_{i-1,j}$ and $C_{up}$ can be described as follows:
\begin{equation}
  \label{eq7}
  C_{up} = V_{i,j} \oplus V_{i-1,j}
\end{equation}

The other directions are similar to Eq.(\ref{eq7}). We store all bitwise xor computation results in the four directions as four redundant tables respectively. Then
the fault-correction mechanism can correct $V_{i,j}$ according to Eq.(\ref{eq8}): 
\begin{equation}
  \label{eq8}
  \left\{
  \begin{aligned}
  &V^1_{i,j} = V_{i-1,j} \oplus C_{up}\\
  &V^2_{i,j} = V_{i+1,j} \oplus C_{down}\\  
  &V^3_{i,j} = V_{i,j-1} \oplus C_{left}\\  
  &V^4_{i,j} = V_{i,j+1} \oplus C_{right}
  \end{aligned}
  \right.
\end{equation}

The correction results in Eq.(\ref{eq8}) are $V_{i,j}$ with different superscripts, as there are probably some faulty elements in the S-box that lead to faults in the correction results.
If there is one faulty element in the S-box, it alters one of the four correction results $V^1_{i,j}$, $V^2_{i,j}$, $V^3_{i,j}$ and $V^4_{i,j}$. 
However, the discriminator can select the majority of the correction results as the final output $V_{i,j}$.

Note that in the horizontal direction,  $C_{left}$ of $V_{i,j}$ is identical with  $C_{right}$ of $V_{i,j-1}$. In the vertical direction,
 $C_{up}$ of $V_{i,j}$ is identical with  $C_{down}$ of $V_{i-1,j}$ as well. Based on this observation, we can condense the four redundant tables in the four directions to two redundant tables in the horizontal and vertical directions. When we correct the adjacent elements, we only need to change the sequential order of elements in the redundant tables.

\begin{figure}[tb]
	\centering	
	\subfloat[The element $82$ with \\its adjacencies]
	{
		\begin{minipage}[t]{0.25\textwidth}
			\centering   
			\includegraphics[width=0.7\textwidth]{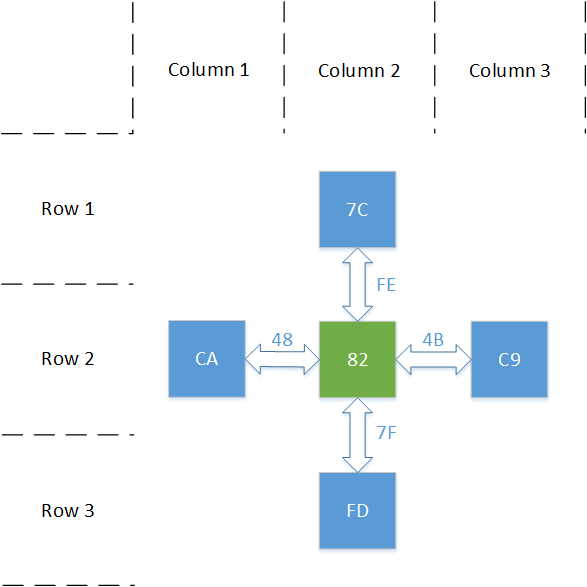} 
			\label{fig4a}
		\end{minipage}
	}
	\subfloat[The element $63$ with \\its adjacencies]
	{
		\begin{minipage}[t]{0.25\textwidth}
			\centering 
			\includegraphics[width=0.7\textwidth]{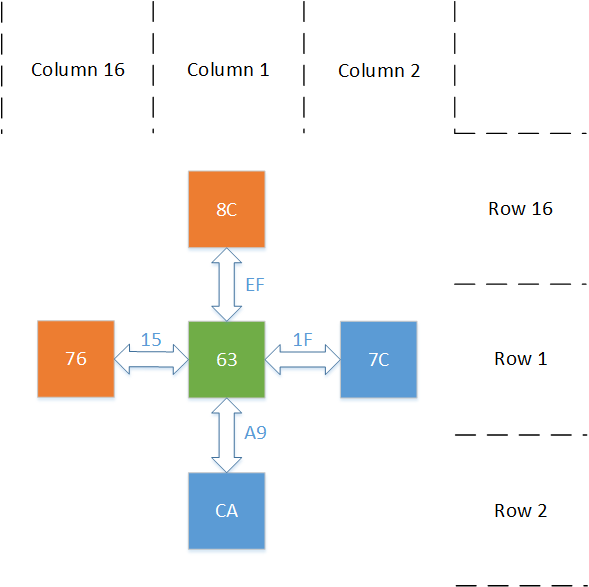} 
			\label{fig4b} 
		\end{minipage}
	}	
	\caption{Toy example of the elements in AES S-box with their adjacencies.}
	\label{fig4}
\end{figure}

Fig.  \ref{fig4} shows two cases in AES S-box. 
The most common case is shown in Fig.  \ref{fig4a}.
The green element 82 has four blue adjacencies. Then we can obtain four bitwise xor computation results to build the redundant tables.
Another special case for the peripheral elements is shown in Fig.  \ref{fig4b}.
The green element 63 is in row 1, column 1. It only has two blue adjacencies in the right and down directions.  To solve this problem,
we connect the S-box table end to end in the horizontal and vertical directions. Row 1 is on the bottom of row 16, and  column 1 is on the right side of column 16.
As shown in Fig.  \ref{fig4b}, two orange elements are used to make up the adjacencies of the green element 63.

\subsection{Time Cost Analysis}
\label{sec3.5}
In this subsection, we analyse the time cost on the basis of our algorithm. In addition, we compare our algorithm with the other two FA countermeasures 
which are introduced in Sect. \ref{sec2}.  For simplicity, we only consider the case that the operations are performed serially. 

Recall the process of AES encryption. Each transformation round consists of four different operations.
And an extra KeyExpansions operation is used to generate the round key from the master key. Let $T_{Add}$, $T_{Sub}$, $T_{Shift}$,
$T_{Mix}$ and $T_{Key}$ denote the time cost of the five different operations (AddRoundKey, SubBytes, ShiftRows, MixColumns and KeyExpansions) respectively. 
Let $T_{Ori}$ denote the time cost of the original AES encryption. We can easily deduce the
following:
\begin{equation}
  \label{eq9}
  T_{Ori} = 11T_{Add} + 10T_{Sub} + 10T_{Shift} + 9T_{Mix} +10T_{Key}
\end{equation}

Focus on our algorithm. Compared with the original AES algorithm, the detection mechanism and the fault-correct mechanism are the additional cost.
Let $T_{Detect}$ and $T_{Correct}$ denote the time cost of the two mechanisms respectively.
In fact, the essence of the detection mechanism is performing the SubBytes operation 20 times, so $T_{Detect}$ can be described as:
\begin{equation}
  \label{eq10}
  T_{Detect} = 20T_{Sub}
\end{equation}

On the other hand, the size of $T_{Correct}$ 
depends on the correction range of the fault-correction mechanism.  If the fault-correction mechanism only corrects the fault element in the S-box, four bitwise xor computations will be 
performed. If the correction range of the fault-correction mechanism is the full S-box, the number of bitwise xor computations is equal to $4 \times 256 = 1024$. 
As the AddRoundKey operation contains 16 bitwise xor computations, $T_{Correct}$ can be denoted as:
\begin{equation}
  \label{eq11}
  0.25T_{Add} \leq T_{Correct} \leq 64T_{Add}
\end{equation}

Let $T_{Algo}$ denote the time cost of our algorithm. It can be easily computed as the sum of $T_{Ori}$, $T_{Detect}$ and $T_{Correct}$. 
So $T_{Algo}$ can be deduced as follows: 
\begin{equation}
  \label{eq12}
  \begin{aligned}
    &11.25T_{Add} + 30T_{Sub} + 10T_{Shift} + 9T_{Mix} +10T_{Key} \leq T_{Algo}\quad \\
    &\quad \leq 75T_{Add} + 30T_{Sub} + 10T_{Shift} + 9T_{Mix} +10T_{Key}
  \end{aligned}
\end{equation}

As mentioned in Sect. \ref{sec2}, the two FA countermeasures --- DMR and BS both need two modules to perform
AES encryption independently. So the time cost of the two FA countermeasures is double the original AES. Let $T_{DMR}$ and $T_{BS}$ denote 
the time cost of DMR and BS respectively, we can denote as:
\begin{equation}
  \label{eq13}
  \begin{aligned}
  T_{DMR} &= T_{BS} = 2T_{Ori}
  \\ &= 22T_{Add} + 20T_{Sub} + 20T_{Shift} + 18T_{Mix} +20T_{Key}
  \end{aligned}
\end{equation}

For the sake of comparison, let $T_{Unit}$ denote the unit of the time cost.  We assume that $T_{Add}$ is equal to $T_{Unit}$. Actually, in \cite{mangard2008power}, the authors pointed out that there are significant differences among the time cost of different operations.
$T_{Add}$, $T_{Sub}$ and $T_{Shift}$ are approximately equal.
$T_{Mix}$ is approximately equal to double the sum of $T_{Add}$, $T_{Sub}$ and $T_{Shift}$. 
And $T_{Key}$ is approximately equal to a quarter of $T_{mix}$.
To sum up, $T_{Mix}$ is the maximal time computation and $T_{Sub}$ can be considered as the minimal time cost in AES encryption.
Then the time cost of our algorithm and the two other FA countermeasures can be written as:
\begin{equation}
  \label{eq14}
  \begin{aligned}
    120.25 T_{Unit} \leq& T_{Algo} \leq 184 T_{Unit}\\
    T_{DMR} =& T_{BS} = 200 T_{Unit}
  \end{aligned}
\end{equation} 

In conclusion, even in the most complex case, our algorithm has less time cost than the two other FA countermeasures.
Moreover, in the best case, the time cost of our algorithm is nearly 40\% lower than the two other FA countermeasures.

\section{Improved Algorithm against Multiple Faults}
\label{sec4}
In this section, we show the performance of our algorithm in the Multiple-Bytes-Faults scenario. Unlike the Single-Byte-Fault scenario,
depending on the positions of the faulty elements, the fault model can be divided into the best case, the average case and the worst case.
In different cases, the performance of our algorithm is distinct.
However, by repeating the process of correction, our algorithm has the potential to correct the faulty elements in the S-box
even in the worst case. 

Each case is discussed in a separate subsection. In each subsection, we describe the fault positions first. Then we discuss the performance of our algorithm and update 
the security mechanism to correct the faulty elements in the S-box.

\subsection{The Best Case}

\begin{figure*}[tb]
	\centering	
	\subfloat[The best case]
	{
		\begin{minipage}[t]{0.33\textwidth}
			\centering   
			\includegraphics[width=0.6\textwidth]{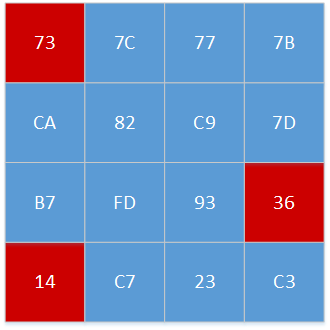} 
		\end{minipage}
		\label{fig5a}}
	\subfloat[The average case]
	{
		\begin{minipage}[t]{0.33\textwidth}
			\centering 
			\includegraphics[width=0.6\textwidth]{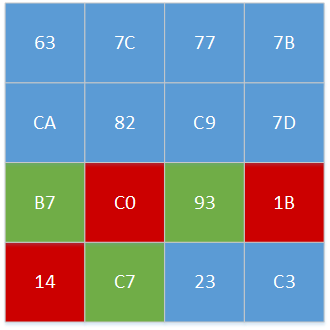} 
		\end{minipage}
		\label{fig5b}}
	\subfloat[The worst case]
	{
		\begin{minipage}[t]{0.33\textwidth}
			\centering 
			\includegraphics[width=0.6\textwidth]{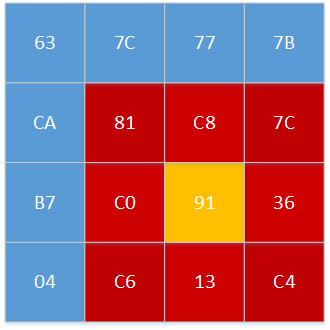} 
		\end{minipage}
		\label{fig5c}}
	\caption{The positions of the faulty elements in the different cases.}
	\label{fig5}
\end{figure*}

  $\emph{\textbf{Fault Positions.}}$ In the best case, each element in the S-box has at most one faulty adjacency.
   As shown in Fig.  \ref{fig5a}, the positions of the red faulty elements are random and scattered.

 $\emph{\textbf{Performance.}}$ According to the description in subsection \ref{sec3.4}, one faulty element only leads to one faulty correction result.
  As each element has four adjacencies, the majority of the correction results are not changed. The discriminator can keep the output correct in the end. 
  
  $\emph{\textbf{Update.}}$ As the fault-correction mechanism is not affected by the multiple faults in the best case. 
  We do not need to update the security mechanism.

\subsection{The Average Case}
  $\emph{\textbf{Fault Positions.}}$ In the average case,  each element in the S-box has at most two faulty adjacencies. As shown in Fig.  \ref{fig5b}, the blue correct elements have at most one faulty adjacency as before.
   However, each green element has two faulty adjacencies.

 $\emph{\textbf{Performance.}}$ It is a challenge for our algorithm to correct the green elements. Half of the correction results of the green elements are still correct.
   However, the other faulty half may impact the output in the end. Focus on the green element B7 or C7 in Fig.  \ref{fig5b}.
   The faulty correction results of the element are different. Then the output of discriminator  keeps correct because the majority of the correction results are correct.
   
   Worse yet, focus on the green element 93. The faulty correction results of the element are the same. The discriminator can not distinguish which output is correct.
  
   $\emph{\textbf{Update.}}$ There are two ways to update our security mechanism. A straightforward way is to add the adjacencies of each element. In the broad sense, considering the diagonal directions, each element has eight adjacencies at most.
   More adjacencies supply more redundancies.  The two faulty adjacencies can not impact the output, since the faulty correction results are the minority in all cases. 
   
   The other way is  repeating the process of correction. Actually, repeating the process of correction twice is enough in the average case. In the first correction, the adjacencies of the green elements are corrected. 
   After the first correction, the average case is the same as the best case. Then in the second correction, the green elements are corrected with the correct adjacencies. Finally, the discriminator keeps the output correct.

\subsection{The Worst Case}
  $\emph{\textbf{Fault Positions.}}$ In the worst case, the faulty elements gather in a large area of the S-box. As shown in Fig.  \ref{fig5c}, the red elements and the yellow element are faulty at the same time.
   Actually, this type of case might seem unlikely. However, if the adversary increases the pulse-width of the laser in the fault injection phase, a large area of elements in the S-box can be affected together.
 
   $\emph{\textbf{Performance.}}$ For the red elements which have two faulty adjacencies, the 
 performance analysis is identical to the average case. 
  Nevertheless, for the yellow element, all its adjacencies and itself are faulty. So that there are not any more redundancies that can be used in the fault-correction mechanism.
  There is no way to correct the faulty element in one correction.
  
  $\emph{\textbf{Update.}}$ Even adding the adjacencies is ineffective. We can still update our security mechanism by repeating the process of correction. The effect of this method depends on the number and position of the fault elements. However, too many faults might be not beneficial to the adversary, we discuss this situation in Sect. \ref{sec6}.

\section{Practical Implementation}
\label{sec5}
To ensure the practicability and   repeatability of our algorithm, we conduct experiments with AES in both software and hardware implementations.
In this section, we introduce the experimental targets in the first place. Then we show the experimental details and results.
Specially, we improve our algorithm using the pre-correction mode in order to reduce the cost of the hardware implementation.

\subsection{Target Devices}
We use two different targets in our experiments. 
The software implementation of AES is placed in $\mu$C, and the hardware implementation of AES is placed in FPGA.
The $\mu$C is the STM32F103 development board  with an  ARM Cortex-M3 core. 
The FPGA is the DE2-115 development board with a Cyclone IV 4CE115F29 core. 
The two targets are common and widely-available so that other researchers can reproduce our work easily.
The details of the targets hardware can be seen in Table \ref{tab1}.

\begin{table}[tb]
  \centering
  \caption{Overview of target devices.}
  \label{tab1}
  \begin{tabular}{ccccc}
  \hline
  Device    & Core                                     & Clock Speed & SRAM Size \\ \hline
  STM32F103 & ARM Cortex-M3                            & 72MHz       & 96KB      \\
  DE2-115   & \multicolumn{1}{l}{Cyclone IV 4CE115F29} & 50MHz       & 2MB       \\ \hline
  \end{tabular}
\end{table}

\subsection{Experiments in the Software AES Implementations}
\label{sec5.2}
On the STM32F103 development board, we run four kinds of AES implementation in both Single-Byte-Fault and Multiple-Bytes-Faults scenarios. All of them are written in C.
The first kind of AES implementation called Ori-AES is the original AES without any countermeasures against FA.
Then the DMR-AES is the AES  which is protected by the REDMR countermeasure with ZCO defense.
The third AES implementation named BS-AES is the AES which is protected by the BS countermeasure using the special ShiftRows operation in the last round.
In the end, the D\&C-AES  refers to the AES which is protected by our fast-detection and fault-correction algorithm.

\begin{table}[tb]
  \centering
  \caption{ Comparison of four AES software implementations.}
  \label{tab2}
  \begin{tabular}{cccc}
  \hline
  Implementation & FI detection & Countermeasure     & Time Cost ($T_{Unit}$)    \\ \hline
  Ori-AES        & No           & N/A                & 100      \\
  DMR-AES        & Yes          & DMR-ZCO            & 200     \\
  BS-AES         & No           & Bytes scrambling   & 200     \\
  D\&C-AES       & Yes          & Our algorithm      & 120.25 $\sim$ 184 \\ \hline
  \end{tabular}
\end{table}

We use the same description $T_{Unit}$ as in subsection \ref{sec3.5}  to denote the unit of the time cost.
As shown in Table \ref{tab2}, the Ori-AES is the simplest implementation and costs 100 $T_{Unit}$.
The DMR-AES and the BS-AES have double cost than the Ori-AES because they need two independent modules to work together. 
Compared with the others, the cost of our D\&C-AES is 120.25 $\sim$ 184 $T_{Unit}$. The extra cost of the D\&C-AES  only involves the detection mechanism and the correction mechanism. 
We demonstrated above that the two extra mechanisms have lower cost than an extra AES module in the DMR-AES  and the BS-AES.

\begin{figure*}[tb]
	\centering 
	  \begin{minipage}[b]{0.35\textwidth}
		  \subfloat[Ori-AES]{
		  \centering   
		  \includegraphics[width=0.8\textwidth]{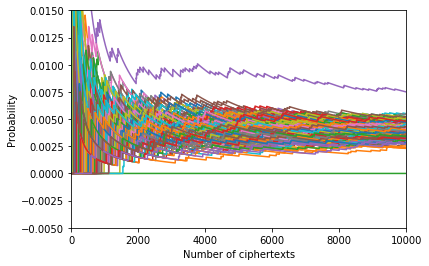} 
  
	\label{fig6a}
  }
  \end{minipage}
  \begin{minipage}[b]{0.35\textwidth}
	\subfloat[DMR-AES]{
  
		  \centering   
		  \includegraphics[width=0.8\textwidth]{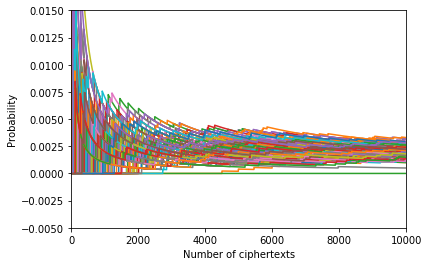} 
  
	\label{fig6b}
  }
  \end{minipage}
  
  \begin{minipage}[b]{0.35\textwidth}
	  \subfloat[BS-AES]{
	  \centering   
	  \includegraphics[width=0.8\textwidth]{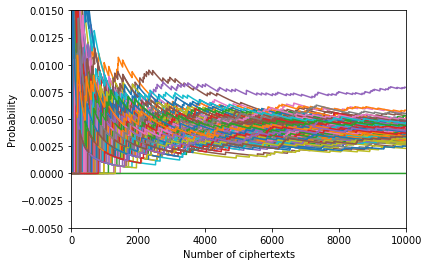} 
  
  \label{fig6c}
  }
  \end{minipage}
  \begin{minipage}[b]{0.35\textwidth}
  \subfloat[D\&C-AES]{
  
	  \centering   
	  \includegraphics[width=0.8\textwidth]{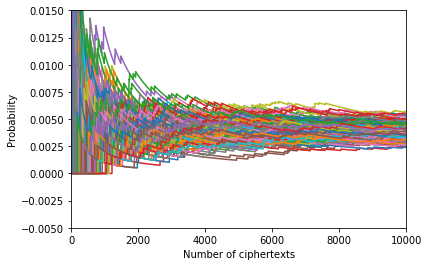} 
  
  \label{fig6d}
  }
  \end{minipage}
	\caption{The distributions of ciphertext values of four AES software implementations in the Single-Byte-Fault scenario.}
	\label{fig6}
  \end{figure*}

\begin{figure*}[tb]
	\centering 
	  \begin{minipage}[b]{0.35\textwidth}
		  \subfloat[Ori-AES]{
		  \centering   
		  \includegraphics[width=0.8\textwidth]{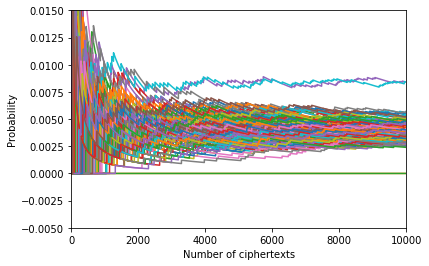} 
  
	\label{fig7a}
  }
  \end{minipage}
  \begin{minipage}[b]{0.35\textwidth}
	\subfloat[DMR-AES]{
  
		  \centering   
		  \includegraphics[width=0.8\textwidth]{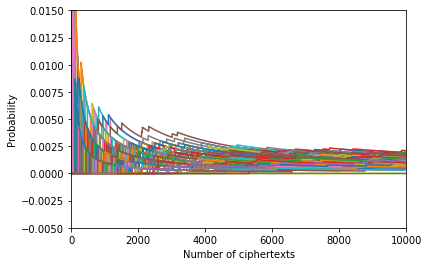} 
  
	\label{fig7b}
  }
  \end{minipage}
  
  \begin{minipage}[b]{0.35\textwidth}
	  \subfloat[BS-AES]{
	  \centering   
	  \includegraphics[width=0.8\textwidth]{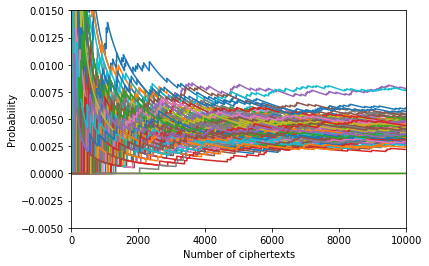} 
  
  \label{fig7c}
  }
  \end{minipage}
  \begin{minipage}[b]{0.35\textwidth}
  \subfloat[D\&C-AES]{
  
	  \centering   
	  \includegraphics[width=0.8\textwidth]{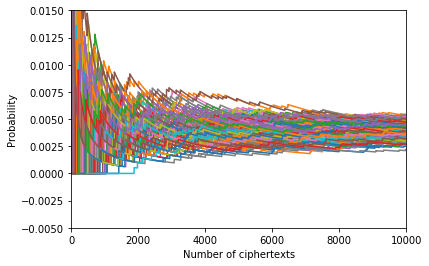} 
  
  \label{fig7d}
  }
  \end{minipage}
	\caption{The distributions of ciphertext values of four AES software implementations in the Multiple-Bytes-Faults scenario with two random faults.}
	\label{fig7}
  \end{figure*}

The experimental details are as follows. For each implementation, we instruct the device to perform AES encryption in both Single-Byte-Fault and Multiple-Bytes-Faults scenarios with a 16 bytes random plaintext.
We repeat the encryption 10000 times and collect the ciphertexts. In the Multiple-Bytes-Faults scenario, we randomly choose two elements in AES S-box to be corrupted to two random values.
The experimental results are shown in Fig.  \ref{fig6} and Fig.  \ref{fig7}.
The probability for each ciphertext value is plotted as one curve, which is calculated as the
counts of appearances for that specific value divided by the number of ciphertexts already
used in the experiments. For each implementation, we repeat
PFA 1000 times on different datasets. The final attack results with the minimum number of ciphertexts are shown in Table \ref{tab3}.

\begin{table}[tb]
  \centering
  \caption{The minimum number of ciphertexts to recover the key of four AES software implementations in the Single-Byte-Fault scenario.}
  \label{tab3}
  \begin{tabular}{cc}
  \hline
  Implementation  & Ciphertext Numbers \\ \hline
  Ori-AES         & 913                \\
  DMR-AES         & 2007               \\
  BS-AES          & 1162               \\
  D\&C-AES        & N/A$^{*}$          \\ \hline
  \end{tabular}\\
  \footnotesize{$*$ none of 16 key bytes were recovered during our experiments}
\end{table}

We consider the Single-Byte-Fault scenario first. For the Ori-AES in Fig.  \ref{fig6a}, we can observe that two curves do not converge on the probability $2^{-8}$.
As our analysis in subsection \ref{sec2.3}, the higher purple curve represents the ciphertext value $c^{max}_j$ whose 
probability is approximately $2 \times 2^{-8}$. And the lower green curve represents the ciphertext value $c^{min}_j$
whose probability is always zero. This confirms that the original AES is vulnerable to PFA. 
Table \ref{tab3} shows that the minimum number of ciphertexts to recover the key of Ori-AES implementation is 913.

Fig.  \ref{fig6b} shows the effect of the DMR countermeasure against PFA. 
Owing to the ZCO defense, each curve converges to a slightly smaller probability than  $2^{-8}$. 
The improvement is that the curve of $c^{max}_j$ converges to the majority anew.
However, the curve of $c^{min}_j$ still remains. 
The adversary can recover the key with Eq.(\ref{eq4}) using 2007 ciphertexts at least.

The BS-AES implementation has a poor performance in our experiments. 
As Fig.  \ref{fig6c} shows, this classical FA countermeasure has little effect on preventing PFA. 
It has a similar distribution probability of ciphertext values as the Ori-AES implementation.
Actually, the minimum number of ciphertexts to recover the key is 1162, which is slightly more than the Ori-AES implementation.
This is because that BS is directed at the transient fault and should be placed after the fault disappears. 
In our experiments, the fault is persistent and does not disappear until the device reboots. As a result, BS is unable to protect the ciphertext effectively.

The experimental result of our algorithm is shown in Fig.  \ref{fig6d}. Unlike the other three AES software implementations,
all curves of our D\&C-AES implementation  converge on the probability $2^{-8}$. Table \ref{tab3} shows that PFA can not recover the key with all ten thousand ciphertexts, which is well in excess of the requirements of the attack in \cite{zhang2018persistent}.
It demonstrates that our algorithm can completely prevent PFA in the Single-Byte-Fault scenario.

As the comparison with the Single-Byte-Fault scenario, the experimental results in the Multiple-Bytes-Faults scenario are shown in Fig.  \ref{fig7}.
Different from the Single-Byte-Fault scenario, two curves are higher than the majority in Fig.  \ref{fig7a} and Fig.  \ref{fig7c}.
The reason is that two faults in the S-box lead to the abnormal probabilities of four ciphertext values (two probabilities rise to $2 \times 2^{-8}$ and two probabilities drop to 0).
For the first three implementations, if the adversary wants to recover the key as before, 
he needs to expand PFA to the penultimate round of AES.

\begin{figure}[tb]
  \centering
  \includegraphics[width=6cm]{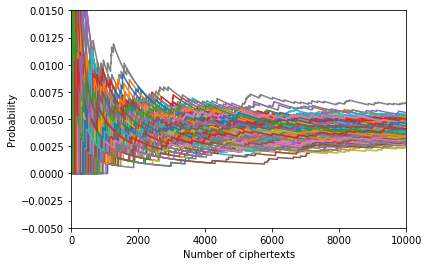}
  \caption{The distributions of ciphertext values of the D\&C-AES software implementations in the worst case with eight random faults.}
  \label{fig8}
\end{figure}

For our D\&C-AES implementation, the result in Fig.  \ref{fig7d} is as good as before. 
Note that there are only two random  faults in our experiments, which is corresponding to the best case or the average case in Sect. \ref{sec4}.
With the increase of the number of faults, our D\&C-AES implementation can not ensure the convergence for every curve.
Fig.  \ref{fig8} shows the distributions of ciphertext values in the worst case with eight random faults. 
It is observed that the performance of our D\&C-AES implementation is far worse than before.
However, it might be hard for adversary to recover the key using PFA as well. The relationship between the performance of PFA and the number of faults is discussed in Sect. \ref{sec6}.

\subsection{Experiments in the Hardware AES Implementations}
The experimental details in the hardware AES implementations are quite similar to that in the software AES implementations. And the experimental results are consistent with
those shown in Fig. \ref{fig6} and Fig. \ref{fig7}.
 The Ori-AES and the BS-AES are vulnerable to PFA. The DMR-AES with ZCO defense can prevent PFA which is based on $c^{max}_j$. However, the adversary can still recover the key using $c^{min}_j$. Compared with the other three AES implementations, our D\&C-AES has the best effect and can prevent PFA effectively.

However, unlike $\mu$C, FPGA requires the designer to implement the design not at the algorithm level but at the circuit level.
The cost of the area and the internal resources must be considered. 

In subsection \ref{sec5.2}, our D\&C AES implementation puts the detection mechanism and the correction mechanism before the encryption.
It is reasonable in $\mu$C because of its serial nature. However, compared with the computation, implementing the circulation is more challenging  and  resource-intensive in FPGA. As there are a mount of circulations in the detection mechanism, 
 we change our algorithm from the D\&C mode to a simplified pre-correction mode to reduce the cost. 
In the pre-correction mode, we remove the detection mechanism and integrate the correction mechanism with the SubBytes operation.
The function of S-box table is replaced with the horizontal and vertical redundant tables which are used in the correction mechanism. 
The output of the correction mechanism is used as the output of the SubBytes operation. 

\begin{figure}[tb]
	\centering	
	\subfloat[D\&C implementation]
	{
		\begin{minipage}[t]{0.23\textwidth}
			\centering   
			\includegraphics[width=1\textwidth]{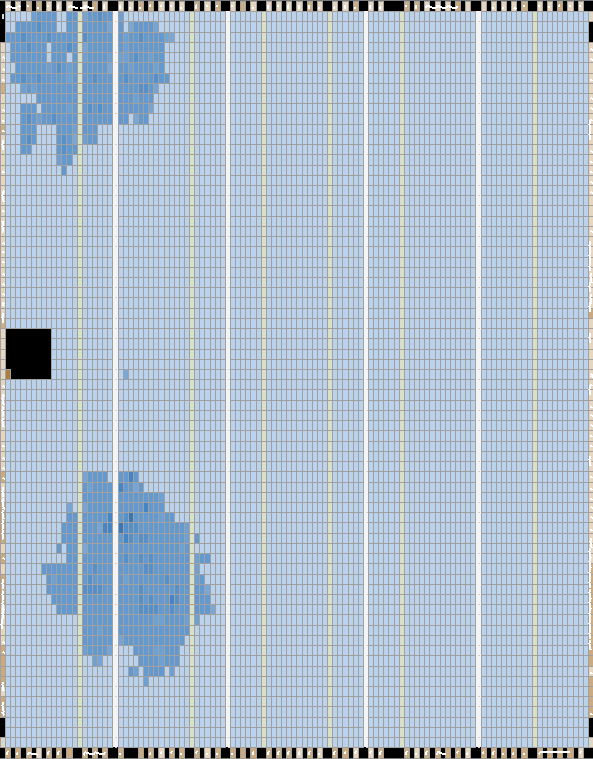} 
		\end{minipage}
		\label{fig9a}
	}
	\subfloat[Pre-correction implementatio]
	{
		\begin{minipage}[t]{0.23\textwidth}
			\centering 
			\includegraphics[width=1\textwidth]{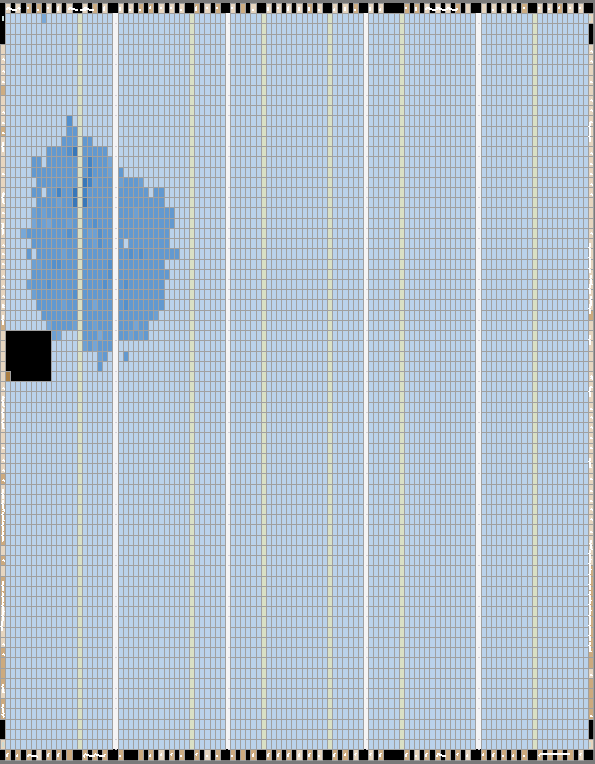} 
		\end{minipage}
		\label{fig9b}
	}	
	\caption{The area cost of the D\&C and the pre-correction implementation.}
	\label{fig9}
\end{figure}

The area cost of the D\&C and the pre-correction implementations is shown in Fig. \ref{fig9}. 
Table \ref{tab4} shows the cost of the internal resources of those two implementations. 
The cost of the pre-correction implementation is about half of the D\&C implementation. 
This  can be boiled down to that the pre-correction implementation reduces the cost of the detection mechanism and the S-box table.

\begin{table}
  \centering
  \caption{The internal resources cost of the two implementations.}
  \label{tab4}
  \begin{tabular}{ccc}
  \hline
  Implementation     & LUTs  & Registers \\ \hline
  D\&C           & 10505 & 403       \\
  Pre-correction & 6438  & 268       \\ \hline
  \end{tabular}
\end{table}

\section{Discussions}
We demonstrate the effectiveness of our algorithm in both software and hardware implementations in the previous section.
In this section, some problems are discussed besides our algorithm itself. We discuss the relationship between the performance of PFA and the number of faults firstly.
Then, we make a comparison between our algorithm and TMR.
Finally, the expansibility of our algorithm is open to discussion. 

\label{sec6}

\subsection{PFA with Multiple Faults}
\label{sec4.4}
In Sect. \ref{sec4}, we find that with the increase of the faulty elements, it is increasingly challenging for our algorithm to correct the fault. 
However, the adversary is not necessarily benefited from the Multiple-Bytes-Faults scenario.
 In \cite{zhang2018persistent}, the authors pointed out that the multiple faults increase the number of ciphertexts required in PFA. If there are $\lambda$ faults in the S-box,
 the remaining key candidates can be at most reduced to $ 16 \times log_{2}\lambda$.
The adversary has to try brute force of the remaining key candidates.
Therefore, if there are too many faulty elements in AES S-box, it is also a challenge for the adversary.

\subsection{Comparing Our Algorithm with TMR}
In subsection \ref{sec3.4}, our fast-detection mechanism corrects faults with the majority voting calculation. This design is similar to TMR, which is another well-known FA countermeasure. However,  using two redundant tables, our algorithm can provide four times redundancy but TMR can only provide double redundancy. Besides, since TMR uses more redundant modules than DMR, the time cost of TMR is higher than DMR as well. As mentioned in subsection \ref{sec3.5}, our algorithm has  less  time  cost  than DMR. Thus, our algorithm is more effective than TMR.

\subsection{Extending Algorithm to Prevent Other FA Methods}
Since our algorithm prevents the attacker to collect 
faulty ciphertexts, those FA methods which require faulty ciphertexts are ineffective.
Take DFA as an example. The adversary compares the differences between  correct ciphertexts and 
 faulty ciphertexts to recover the key. In \cite{jeong2013new}, the authors proposed a new DFA on block cipher with S-box.
Their fault model assumed the adversary has the capability to inject a Single-Byte-Fault into S-box and collect  correct/faulty ciphertexts.
This assumption does not make sense with our algorithm because all  ciphertexts are correct.

\section{Conclusion}
\label{sec7}
In this paper, we propose a fast-detection and fault-correction algorithm against PFA.
The strength of our algorithm is that it can completely prevent PFA by correcting the S-box element fault in the Single-Byte-Fault scenario. Further, in the Multiple-Bytes-Faults scenario,
our algorithm also has good performance. The experimental results in both software and hardware implementations
ensure the practicability and repeatability of our algorithm. Compared with the classical FA countermeasures like DMR and BS, 
our algorithm is more effective and has less time cost.

Our correction mechanism can be easily transplanted to the encryption algorithms based on \emph{Substitution-Permutation Network} (SPN),
 such as LED or PRESENT. 
However, there are also some encryption algorithms, such as RSA or ECC, with different structures. It is a challenging future work to design the correction mechanism for those encryption algorithms.

\section*{Acknowledgment}

The authors would like to thank Information Science Laboratory Center of USTC for the hardware/software services. This work was supported by National Natural
Science Foundation of China (Nos. 61632013, 61972370 and 62002335), and Fundamental Research Funds for Central Universities in China (No. WK3480000007).

\bibliographystyle{IEEEtran}
\bibliography{IEEEabrv,bibfile}

\end{document}